\documentclass[sigconf]{acmart}

\acmConference[ICPC 2022]{The 30th International Conference on Program Comprehension}{May 21–22, 2022}{Pittsburgh, PA, USA}

\usepackage{multirow, tabularx, balance, stfloats}

\usepackage{graphicx}
\usepackage{caption}
\usepackage{subcaption}

\usepackage{tcolorbox}

\copyrightyear{2022}
\acmYear{2022}
\setcopyright{acmcopyright}\acmConference[ICPC '22]{30th International Conference on Program Comprehension}{May 16--17, 2022}{Virtual Event, USA}
\acmBooktitle{30th International Conference on Program Comprehension (ICPC '22), May 16--17, 2022, Virtual Event, USA}
\acmPrice{15.00}
\begin{document}

\title{On the Transferability of Pre-trained Language Models for Low-Resource Programming Languages}

% author names and affiliations
% use a multiple column layout for up to three different
% affiliations
\author{Fuxiang Chen}
\affiliation{
    \institution{University of British Columbia}
    \country{Canada}
}
\email{fuxiang.chen@ubc.ca}
% \and
\author{Fatemeh H. Fard}
\affiliation{
    \institution{University of British Columbia}
    \country{Canada}
}
\email{fatemeh.fard@ubc.ca}
% \and
\author{David Lo}
\affiliation{
    \institution{Singapore Management University}
    \country{Singapore}
}
\email{davidlo@smu.edu.sg}
% \and
\author{Timofey Bryksin}
\affiliation{
    \institution{JetBrains Research}
    \country{Republic of Cyprus}
}
\email{timofey.bryksin@jetbrains.com}

\begin{abstract}
% Pre-trained Language Models (PLM), when fine-tuned on natural language understanding tasks such as Question and Answering, 
%, trained from a large corpus of natural language text i.e., Wikipedia content, 
% have consistently achieved the state-of-the-art results. 
% The Software Engineering (SE) 
% research 
% community has shown increasing interests in the applicability of Pre-trained Language Model (PLM) in SE-related tasks. 
% Pre-trained Language Models (PLM) such as CodeBERT and GraphCodeBERT, when trained on a large corpus of code, have recently displayed promising results in Software Engineering (SE) downstream tasks.
% such as 
% the downstream tasks: 
% code summarization.
% A PLM is most useful if it can be leveraged to improve the performance on code corpora written in low-resource programming languages, where training data is limited. In this work, our focus is on studying the impact of PLMs on a low-resource programming language corpus --- specifically, we choose Ruby as the study subject. 
% In contrast to NLP, 
% Previous work has reported high effectiveness on using PLMs 
% but a deeper analysis on the PLM's capability on SE-related tasks is limited. 

A recent study by Ahmed and Devanbu reported that using a corpus of code written in multilingual datasets to \emph{fine-tune} multilingual Pre-trained Language Models (PLMs) achieves higher performance as opposed to using a corpus of code written in just one programming language.
However, no analysis was made with respect to fine-tuning monolingual PLMs. Furthermore,
some programming languages are inherently different and code written in one language usually cannot be interchanged with the others, i.e., Ruby and Java code possess very different structure.
% In order to leverage the learned knowledge from PLM effectively, 
To better understand how monolingual and multilingual PLMs affect different programming languages,
% it is important to understand the behavior of PLM when applied to SE-related tasks. 
%However, as there is a recent attention to PTLMs in SE, the studies to evaluate their capabilities are limited. 
% In this work, 
we investigate 
% the following: 
1) the performance of 
% using  
% monolingual and multilingual 
PLMs 
% fine-tuned on multilingual data 
on Ruby for two popular Software Engineering tasks: Code Summarization and Code Search,
2) the strategy (to select programming languages) that works well on fine-tuning multilingual PLMs for Ruby, and 
3) the performance of the fine-tuned PLMs on Ruby given different code lengths.
% --- here, we bin the Ruby code based on its number of tokens; understanding the performance on different code lengths will enable developers to make more informed decision on the use of PLMs based on their code.
% 1) Transferability on the 
% % zero-shot settings for 
% PLM pre-trained with mainstream programming languages, Java and Python, 
% % (i.e.,  no fine-tuning is conducted in the downstream  tasks), 
% 2) Transferability on the PLM pre-trained with low-resource languages, and 3) Transferability on random initialised PLM. 
%, as well as their essential characteristics (such as the required number of epochs) when fine-tuned on SE-related downstream tasks.
%\textcolor{red}{We find that TODO... The results of our research opens new insights about XX for researchers and developers using these models.}
%Over 100 models were trained/fine-tuned during a period of one year. 

In this work, we analyze over a hundred of pre-trained and fine-tuned models. Our results show that 1) multilingual PLMs have a lower Performance-to-Time Ratio (the BLEU, METEOR, or MRR scores over the fine-tuning duration) as compared to monolingual PLMs, 2) our proposed strategy to select target programming languages to fine-tune multilingual PLMs is effective --- it reduces the time to fine-tune yet achieves higher performance in Code Summarization and Code Search tasks, and 3) our proposed strategy consistently shows good performance on different code lengths.
% Adding more dataset of other programming languages during pre-training does not help in the fine-tuning tasks (in fact, it worsens the results), and 2) PLMs that are trained and fine-tuned using low-resource languages have similar performance with PLMs that are trained with multiple languages.
% We note here that training a new PLM requires high computation resources. 
% The findings in this study is important and they can help developers to save time by making more informed decision in using PLMs for their tasks. 
% We will release the code, models, and data upon paper acceptance.
\end{abstract}



%%
%% Keywords. The author(s) should pick words that accurately describe
%% the work being presented. Separate the keywords with commas.

\maketitle

% \begin{IEEEkeywords}
% Neural pre-trained language models, language models for source code, transfer learning
% \end{IEEEkeywords}

% \IEEEpeerreviewmaketitle

\section{Introduction}
\label{Section:Introduction}

%HI! some of the citations to use are CodeBERT, CuBERT, and there are a few others I can add later, when talking about PTLMs for code or the use of DL for other SE tasks. 
% \fuxiang{Thanks! I will try to include those citations where I think is necessary in the papers.}

% \fuxiang{
% Proposed Introduction Story Line:
% \\1) Talk about how NLP PLM are becoming the main stream in NLP Tasks
% \\2) Talk about recent growing interests in the SE Community in applying PLM on SE-related Tasks - Why the growing interests? e.g., one of the reasons is due to the high similarity between natural language text such as English and Programming Language (The naturalness hypothesis)
% \\3) Talk about the continuous improvements of PLM (mainly on NLP PLM - on using different learning objectives). 
% \\4) Talk about why we want to perform this study - e.g, why this is important and applicable to developers, etc *** I think this point needs to be stressed very clearly
% \\5) Talk about the focus of this paper
% \\6) Talk about the results
% }

%Briefly talk about NLP's PLM.
% Recent progress in Natural Language Processing (NLP) has been driven by the adoption of deep neural models requiring large amount of labeled data \cite{beltagy-etal-2019-scibert}. However, collecting and annotating such data can be challenging and expensive \cite{beltagy-etal-2019-scibert}. Several studies have shown that 
% Using 
Unsupervised pre-training of language models on large corpora significantly improves the performance in many downstream tasks \cite{bert,roberta}. In this work, we refer to a Pre-trained Language Model and its plural form as PLM and PLMs, respectively. 
% 
% Different domains \cite{lee2020biobert,naseem2020biomedical,araci2019finbert,liu2020finbert,lu2019vilbert,su2019vl}, including Software Engineering \cite{karampatsis2020scelmo,buratti2020exploring,svyatkovskiy2020intellicode,clement2020pymt5,mastropaolo2021studyingT5,kanade2020learningCuBERT,guo2020graphcodebert,ahmad2021unifiedPre-training,wang2021codet5,roziere2020unsupervisedTransCoder,wang2022syncobert,jiang2021treebert,liu2020multi,feng2020codebert} have leveraged PLMs on their domain-specific tasks. 
% Specifically, these PLMs are trained with domain-specific corpora and/or with new pre-training objective functions. 
% For Software Engineering (SE), in recent years, several PLMs have been released.
% They are either pre-trained on a programming language or a combination of programming languages. 
There have been some attempts to understand how PLMs affect the performance of different downstream tasks empirically \cite{ciniselli2021empirical,austin2021program,mahmud2021code,nie2021evaluation,gros2020code,feng2020codebert}. 
% Separately, Feng et al. conducted a small zero-shot setting experiment on code summarization \cite{feng2020codebert}. Zero-shot setting refers to fine-tuned on a programming language (C\#) not seen during pre-traning \cite{feng2020codebert}. 
% The zero-shot language used is the C\# programming language, which has not been seen by the PLMs. 
% We note that C\# and Java have very similar structure, and one of the programming languages used in pre-training the PLM in Feng et al.'s study is Java. There are mixed results in performing zero-shot code summarization using different PLMs and no further studies have been conducted.

Despite the existing efforts to understand PLMs, there are still many unknowns on the transferability of PLMs for programming languages. \textbf{Firstly}, existing PLMs are trained either on a programming language or on multiple programming languages. Little is known if PLMs trained on a particular programming language yields better performance than a more general PLM that is trained on multiple programming languages. 
A closely related work is by Ahmed and Devanbu \cite{ahmed2021multilingual} that studied the effects of \emph{fine-tuning} for publicly available multilingual pre-trained models, CodeBERT and GraphCodeBERT. However, some programming languages are inherently different, so utilizing a single multilingual model may not always yield the best performance.
\textbf{Secondly}, the zero-shot setting in previous work is lightly studied \cite{feng2020codebert}. For example, other programming languages may have very different structures as compared to the programming languages used to pre-train PLMs. Also, during the pre-training, fine-tuning, and testing processes, different programming languages may be used. To better understand the zero-shot setting, these need to be studied more thoroughly.
\textbf{Thirdly}, the current datasets that are released for training PLMs on Software Engineering (SE) related tasks such as Code Summarization, are dominated by a few languages, mainly Java and Python \cite{husain2019codesearchnet} --- they are known as high-resource programming languages as there is a high volume of code written in them. %languages due to their high volume of availability. 
% Many SE datasets are also largely in Java and Python.  
Other programming languages are often missing or have low number of records in the datasets --- they are known as low-resource programming languages \cite{husain2019codesearchnet}.
In a recent survey conducted by StackOverflow, we observe that although Java and Python are among the popular programming languages used by developers, developers also reported that they are using 36 other programming languages such as Ruby, Kotlin, and Scala 
%developers are using many other programming languages than Java and Python. 
% among the surveyees, 84.2\% of them reported that they are using other programming languages (only 15.8\% of the surveyees reported that they are using Java and Python) 
\cite{so-survey}.
% \fuxiang{To support why other programming languages are also important.}
% \fuxiang{Need to say why this study is on Zero-shot settings because all the experiments are on zero-shot. The training of a PLM on a programming language, fine-tuned on another programming language in a downstream task, and tested on yet another programming language in the same downstream task may look a bit strange. Maybe need to motivate the fine-tuning part to make it clearer.}
%For example, CodeSearchNet only includes 6 programming languages and Go language has XXX records out of YYY total dataset. 
%Thus, it is also important to understand these differences and how to apply them effectively in SE.
Thus, understanding if PLMs pre-trained on high-resource programming languages can be utilized for other programming languages is important.

To bridge the gap in understanding the applicability and the transferability of PLMs 
% (\textit{we do not change the objectives}) 
in SE,
%As the PLMs are findings their way in software engineering and gaining popularity, 
in this study, 
% we investigate the ability of PLMs when pre-trained and fine-tuned on different programming languages. In this work, 
\textbf{we focus on studying the impact of PLMs on a low-resource programming language corpus --- specifically, we choose Ruby as the study subject because it is highly ranked among low-resource languages in the Stack Overflow survey \cite{so-survey} and it is also a commonly used low-resource programming language \cite{husain2019codesearchnet,feng2020codebert,guo2020graphcodebert}}.
We explore five different settings when using PLMs for different downstream tasks:
%to evaluate their applicability and transferability.
%The four settings are: 
\begin{enumerate}
\item Transferability of PLMs pre-trained with a code base written in a single and multiple programming languages: we are interested to know if a monolingual PLM (\textit{a PLM that is pre-trained and fine-tuned on a single programming language}) works better than a multilingual PLM (CodeBERT and GraphCodeBERT).

\item Transferability of PLMs in different zero-shot scenarios: we are interested to know if unseen programming languages can leverage a PLM effectively in downstream tasks. 

\item Efficiency of PLMs: we are interested to know the performance and training time trade-offs among the PLMs.

\item Transferability of PLMs  depending on different code lengths: as developers write code differently, to better understand how a PLM may perform on code of different lengths, in the fourth setting, we are interested to know the performance of PLMs on different code length. 
% Fine-tuning on a suitable programming language is important for good transferability of PLM. 

\item The strategy to select suitable programming languages for fine-tuning: we are interested to explore a strategy to select suitable high-resource programming languages for fine-tuning. 
% We compare the semantic and textual similarities between the high-resource programming language and the target language.
\end{enumerate}
For the dataset, we use CodeSearchNet
% , a dataset published by Github and Microsoft 
\cite{husain2019codesearchnet}
which contains code in six different programming languages, including high-resource and low-resource ones. 
% RoBERTa is used to pre-train the PLMs. 
We note here that the purpose of this study is not to beat the state-of-the-art, but to understand the things mentioned above. Thus, RoBERTa, a strong baseline model used in many PLM related studies and is the basis of the multilingual PLMs (CodeBERT and GraphCodeBERT), is used here for pre-training the PLMs  
% RoBERTa is also  
% of the recent PLM trained on code 
\cite{yang2019xlnet,feng2020codebert,jiao2019tinybert,qiu2020pre,clark2020electra,pradel2018deepbugs,kanade2020learningCuBERT}.
%For 1) and 2), we train and validate the PLM on individual programming language as well as all the programming languages, and 
For all the settings, over 100 PLMs are trained or fine-tuned on different programming languages, and we 
evaluate their performance on two commonly studied SE tasks: Code Summarization and Code Search  \cite{rencos,cocogum,hgnn,ase}.

% , ELECTRA, which is one of the most referred models in NLP for text generation \cite{}. 

%By these different settings, we mainly aim to investigate the ability of the PLMs when fine tuned and tested on different languages. 

Our results show 
%that \textcolor{red}{TODO}.
%These findings can open new avenues for researchers to study different techniques for the transferability of knowledge among different programming languages and for practitioners when using PLMs. 
several interesting phenomena: 1) For Code Summarization, PLMs fine-tuned on the entire multilingual dataset do not yield the best performance but, for Code Search, the best performance is observed on PLMs fine-tuned on the entire multilingual dataset; 2) Monolingual PLMs trained on a combined multilingual dataset have higher Performance-to-Time Ratio (PTR) than multilingual PLMs trained on a multilingual dataset.
The PTR ratio measures the trade-off between the training time to fine-tune a PLM and its performance in downstream tasks
; 3) PLMs fine-tuned on the Python dataset have the best performance in our zero-shot experiments; 4) There are negligible differences in the performance between PLMs tested on a test dataset binned in different code lengths and PLMs tested on the entire test dataset; and 5) Our proposed strategy in selecting a programming language for fine-tuning is effective: it improves the performance over PLMs fine-tuned on the combined multilingual dataset. 
% We note here that training a new PLM will require high computational resources. 
% \textbf{Significance:} 
The findings in this study are important since researchers and practitioners can save time (through our PLM study) and may achieve better performance 
% (our experiments show that our strategy to select a compatible target programming language dataset to fine-tune has better performance than using the combined multilingual dataset) 
by making a more informed decision in using PLM for their tasks. We note that the most recent empricial work on PLMs \cite{ahmed2021multilingual} reported that multilingual PLMs fine-tuned on a combined multilingual dataset perform better in downstream tasks, but our experiments have shown that this might not be the case for all the downstream tasks and we proposed an effective strategy to pick another high-resource language to train on (rather than training on the combined multilingual dataset).

Overall, this paper makes the following contributions:
\begin{itemize}
  \item \textbf{An empirical evaluation on the downstream tasks using monolingual and multilingual PLMs} We perform a detailed quantitative and qualitative evaluation for over a hundred of models on two downstream tasks (Code Summarization and Code Search). 
  \item \textbf{Proposed Strategy to select a suitable PL for fine-tuning PLMs} We proposed an effective strategy to select a suitable programming language for fine-tuning PLMs in Code Summarization and Code Search.
  \item \textbf{Multiple PLMs were trained on different programming languages for two different tasks -- Code Summarization and Code Search.} Training a PLM requires high computational resources. In order to understand the applicability and transferability of PLMs in SE, we have pre-trained and fine-tuned over a hundred PLMs\footnote{https://doi.org/10.20383/102.0563} in different programming languages. Based on our findings, in the downstream tasks, developer can use monolingual PLMs on fine-tuning the combined multilingual datasets which is more time-efficient yet having similar (Code Summarization) or better (Code Search) performance.
%   The pre-trained PLM were used to fine-tune two different tasks: 1) Code Summarization and 2) Code Search. Over 40 models were created during the fine-tuning process.

%   We found several observations such as monolingual PLMs fine-tuned on the combined multilingual datasets have higher performance-to-time ratios than multilingual PLMs fine-tuned on the combined multilingual datasets, and human found negligible differences in the quality of the summaries generated by the different PLMs.
\end{itemize}

The rest of this paper is organized as follows. Related work is surveyed and discussed in Section \ref{Section:Related-Work}. Sections \ref{Section:Method} and \ref{Section:Stratgey} describe our research questions and methods that we employ to answer them, whereas Section \ref{Section:Experimental-Setup} describes the experimental setup. We portray the results for both Code Summarization and Code Search in Section \ref{Section:Results}.
Further discussions on the results are presented in Section \ref{Section:Discussion} and 
various threats are analysed in Section \ref{Section:Threats}. We conclude with directions for future research in Section \ref{Section:Conclusion-and-Future-Works}.

% \section{Background}
% \label{Section:Background}

% \textcolor{red}{Should we include the background section at all?!}
% Discuss the Transformer, PTLMs, and Adapters. 
% How the PTLMs are used and fine tuned for downstream tasks, 
% \fuxiang{Instead of Background, we can have Motivation. I feel that the motivation should be more clear. Proposed Motivation: 1) Creating a new PLM requires huge amount of computational resources and long training time - cite several papers. 2) We perform some simple test and found that using NLP-based PLM on SE-related tasks can achieve similar or better performance than SE-based PLM like CodeBERT (Need to confirm do we have such experiment?). 3) The study of the applicability of NLP-based PLM on SE-related Tasks are limited. Programming Language source code are written in Natural Language Text, and they exhibit multiple similarities in structure and semantics. It is important to perform such study because it will lead to less computation resources requirement and reduced time for developers and researchers. 4) Based on the above reasons, we aim to investigate the transferability of NLP-based PLM and their essential characteristics on SE-related downstream tasks. We describe the research questions in Section \ref{Subsection:Research-Questions}}
 
\section{Related Work}
\label{Section:Related-Work}
Here, we surveyed how different PLMs are used in Software Engineering and discussed the missing gaps.

Kanade et al. use BERT to pre-train a model on Python source code \cite{kanade2019pre}. 
The authors later train a BERT model for source code, known as CuBERT, which is then fine-tuned for classification tasks (e.g., wrong binary operator) and program repair tasks \cite{kanade2020learningCuBERT}.
SCELMO is a PLM based on ELMO that is trained on JavaScript source code for the program repair task \cite{karampatsis2020scelmo}.
Xu et al. \cite{xu2020incorporating} incorporate external knowledge for code generation through pre-training the model with natural language and code pairs and then fine-tune their model for code generation. 
Buratti et al. train BERT on the C language, known as C-BERT, which is used for abstract syntax tree tagging tasks \cite{bui2021Corder-contrastive}. 
PyMT5 uses Transformer to pre-train a model for generating Python methods from docstrings. It also generates code summaries \cite{clement2020pymt5}.
Pre-training a model to represent source code using contrastive learning is proposed in \cite{bui2021Corder-contrastive}. The authors present Corder and use this pre-trained model for Code Retrieval and Code Summarization tasks.
GraphCodeBERT is a PLM that 
% is developed for pre-training on programs and 
uses Transformer as its main architecture. 
The model is tested on Code Search, Code Clone Detection, Code Translation, and Code Refinement tasks \cite{guo2020graphcodebert}. 
CodeBERT is a PLM that uses programming and natural language in the pre-training and combines two training objectives; it is tested on different tasks such as Code Search \cite{feng2020codebert}. 
IntelliCode Compose is a GPT based model that is trained on Python, C\#, JavaScript, and TypeScript for code completion \cite{svyatkovskiy2020intellicode}. 
PLBART is another attempt to build a PLM using the BART architecture \cite{lewis2019bart}. It is trained on Java and Python and is tested on several tasks including Code Summarization, Code Generation, Code Translation, Code Clone Detection, and Program Repair \cite{ahmad2021unifiedPre-training}. 
Text-To-Text Transfer Transformer is another study based on T5 \cite{raffel2019T5} that leverages PLMs to study code related tasks including bug fixing and comment generation \cite{mastropaolo2021studyingT5}. 
Other similar models that leverage PLMs for code related tasks are CodeT5 \cite{wang2021codet5}, Transcoder \cite{roziere2020unsupervisedTransCoder}, SynCoBERT \cite{wang2022syncobert}, TreeBERT \cite{jiang2021treebert}, and a model that uses multi-task learning for pre-training the language model for code completion \cite{liu2020multi}.
All of the above works present a PLM to represent code for different tasks.
% , without extensively presenting evaluations to learn about the PLM's ability for zero shot or transferability to different programming languages. 
% Furthermore, 
Some works focus on pre-training a PLM using a programming language, while others use two to six different programming languages. Little is known if PLMs that are pre-trained on a single programming language yield better results or generalize better on downstream tasks than those that are pre-trained on multiple programming languages.

Some recent works studied deep neural models leveraging PLMs empirically, mostly in their performance on downstream tasks. 
Ciniselli et al. \cite{ciniselli2021empirical} studied how RoBERTa and T5 models affect code completion. The authors concluded that Transformer-based models achieve good results when predicting an entire block of code or when predicting a few masked tokens. 
Mahmud et al. compare three Code Summarization models quantitatively and qualitatively \cite{mahmud2021code}. 
Nie et al. \cite{nie2021evaluation} study the evaluation methodologies for comment generation and method namings, and proposed using time-segmented data for more realistic evaluation, i.e., take train, validation, and test sets from the same year range. 
The metrics, datasets used, and evaluation of comment generation models are also explored in the work of Gros et al. \cite{gros2020code}.
In another study, the limitations of large language models for program synthesis are explored \cite{austin2021program}. 
A more related work by Feng et al. has shown the performance of Code Summarization on C\# code leveraging PLMs in a zero-shot setting \cite{feng2020codebert}. However, the zero-shot experiment is not the focus of their study and it was conducted very briefly. Little is known on the PLM's ability for zero-shot learning or on the transferability to different programming languages.
Although these works explore different PLMs or models developed for a specific code related task, none of them probes the capability of a PLM in different settings
%like we studied in our work
, which are currently missing in the literature. With this study, we intend to start filling this gap. It is important to understand that so that researchers and practitioners can be more productive and effective by making more informed decision in using PLMs for their tasks.

%add the PLM study for sentiment analysis, David Lo

\section{Research Method}
\label{Section:Method}

To better understand the applicability and transferability of the PLMs in SE, we investigate a number of Research Questions (Section \ref{Subsection:Research-Questions}) and describe their study design (Section \ref{Subsection:Study-Design}).

\subsection{Research Questions}
\label{Subsection:Research-Questions}

%Reformulate the RQs and add the rationale (?) and explanation.

% \noindent \textbf{RQ1: Transferability on the PLM  pre-trained with mainstream dataset, Java and Python} 
% %How the PLM perform in zero-shot setting? 
% %How is the transferability on the zero-shot settings for the mainstream languages, Java and Python (\textit{i.e., no fine-tuning is conducted in the downstream tasks})?
% Programming languages share similar traits. We are interested to know if the PLM that is pre-trained with the mainstream dataset, Java and Python, is sufficient for other programming languages (the other programming languages are fine-tuned and evaluated on the PLM). In another setting, we perform zero-shot evaluation by fine-tuning the PLM with the mainstream datasets, and evaluating it with other programming languages.

% \noindent \textbf{RQ1: Does training and fine-tuning on different/multiple programming languages help to improve the quality?} 
% Existing work have used either a single programming language or multiple programming languages to pre-train PLMs. There is no study or comparison done to understand how these PLMs perform in downstream tasks. In this research question, we are interested to know if additional programming languages add merits to pre-training a PLM.

\noindent \textbf{RQ1: Does training and fine-tuning on the individual programming languages improve the performance over multilingual PLMs that are fine-tuned on multilingual datasets?} 
Existing work have used either a single programming language or multiple programming languages (multilingual) to pre-train PLMs. Specially, the multilingual PLMs are CodeBERT and GraphCodeBERT. 
% There is limited study to understand how these PLMs perform in downstream tasks. 
We are interested to know if additional programming languages add merits to pre-training a PLM in different tasks.
\hfill \break

\noindent \textbf{RQ2: Which PLM has the best Performance-to-Time Ratio (PTR)?} 
Although the performance of a PLM is important, the training of a PLM is notoriously known to be computationally expensive. 
We are interested to know the PLMs that have the best trade-off between performance and training time.
\hfill \break

% \noindent \textbf{RQ2}: Are the PLM transferable to a low resource programming language?

% \noindent \textbf{RQ2: Transferability on specialised PLM} 
% %How transferable are the trained PLM on different programming languages settings, including low resource programming languages?
% %How is the transferability on the zero-shot settings for the less popular languages, including low-resourced languages (\textit{i.e., this is similar to the above with the exception that the PLM are tested on the less popular languages})
% We are interested to know if specialised PLM (PLM that is pre-trained and fine-tuned in a programming language) works better than a PLM that is pre-trained with multiple programming languages. This includes low-resource programming languages. If a specialised PLM has better performance, then, less time will be required to train a PLM. 

\noindent \textbf{RQ3: What are the best settings for zero-shot downstream tasks?} 
Feng at al. has conducted a small study on zero-shot Code Summarization using CodeBERT \cite{feng2020codebert}. The unseen programming language used in their experiment was C\# and it has a similar structure to Java, which was used as one of the programming languages to pre-train CodeBERT. There are multiple unknowns in understanding the zero-shot settings on PLMs. For example, if we fine-tune a PLM (pre-trained with programming language A, e.g., PHP) on a programming language B (e.g., Go) and test it on a programming language C (e.g., Ruby), will it have a better performance over a PLM pre-trained and fine-tuned using the same target programming language? We are interested to understand this zero-shot setting. 

\hfill 
% \break

% \noindent \textbf{RQ3: Transferability on random initialised PLM} The downstream tasks require fine-tuning the PLM to create task specific models. Currently, little information is known about the effects of a pre-trained PLM on the downstream task. Here, we investigate a lower bound in the PLM where we initialised its weights randomly.

\noindent \textbf{RQ4: What effect does the PLMs have on different code leng-th?} 
Developers write code in different lengths. However, in many of the existing studies, the performance of the PLM is reported as an average metric score. Thus, it is possible that the reported score may be skewed towards certain code length within the test data --- for example, the test data may contain mostly short code and the reported score may not reflect the true behavior of code in other lengths. We are interested to understand if the PLM has any effect on the length of the code.

\hfill 
% \break

% \noindent \textbf{RQ3: Is training and fine-tuning on high-resource programming languages enough?} Current SE datasets are dominated by a few programming languages, mainly Java and Python. In this RQ, we are interested to know if low-resource programming languages can benefit from PLMs that are trained on high-resource programming languages. We note here that experiments conducted in this RQ may have some overlaps in previous RQs. 

\noindent \textbf{RQ5: How effective is our strategy to decide in advance a language that can work well for a target low-resource language?} Different programming languages have different syntax, and code fragments written in different languages are usually non-interchangeable. For example, code written in Ruby and Java have several differences such as the way data is flowed within code, and we cannot
just rewrite a Ruby fragment with Java constructs and expect it to work.
% copy and paste code from Ruby into Java and expect it to compile without error. 
Here, we propose a strategy to choose in advance programming languages that can work well for fine-tuning multilingual PLMs. We are interested to understand if our proposed strategy is effective.

% How many epochs are required to achieve a good performance when fine tuning the models for a certain task?

% How the PLMs perform for different tasks?
% How each of the PLMs are compared for each of the tasks, ELECTRA vs. RoBERTa?

% What is the effect of the pre-training programming language on the PL of the downstream task?

% How transferable are the pre-trained language models based on the programming language used in the pre-training in zero shot setting?

% Are PTLMs transferrable to low resource programming languages?

% Can PTLMs result in good performance when only fine tuned for one epochs on the downstream task, compared to when they are fully fine tuned (trained) on the downstream task?

% Do the PTLMs have the same effect when applied on different tasks?

% Do the PTLM have the same effect when different models are used?

\subsection{Study Design}
\label{Subsection:Study-Design}

\noindent \textbf{RQ1 Design:} We train multiple PLMs that will be used for fine-tuning on two different widely used downstream tasks: Code Summarization and Code Search  \cite{ahmed2021multilingual,feng2020codebert,guo2020graphcodebert}.
% Specifically, RoBERTa will be used for fine-tuning the Code Summarization task, while ELECTRA will be used for fine-tuning the Wrong Binary Operator Detection task. 
The PLMs are pre-trained on individual programming languages. We use CodeSearchNet, a popular dataset consisting of six programming languages published by Github and Microsoft \cite{husain2019codesearchnet}
% . Data for each programming language is extracted from CodeSearchNet 
to pre-train and fine-tune PLMs.
% We also pre-train a PLM that consists of the entire CodeSearchNet dataset for the multiple programming languages setting. 
% After the pre-training is done, we use different individual programming languages to fine-tune the PLM for the downstream tasks.
We compare \emph{monolingual PLMs} (pre-trained on individual programming languages) fine-tuned on a  \emph{monolingual dataset} with the PLMs fine-tuned on a \emph{multilingual datasets}. Here, the multilingual dataset refers to the combined dataset of all the programming languages. We also compare the best monolingual PLMs fine-tuned on each monolingual dataset with multilingual PLMs (CodeBERT and GraphCodeBERT) fine-tuned on the multilingual dataset which have reported having the best performance in Code Summarization and Code Search \cite{ahmed2021multilingual}.
Additionally, we perform human evaluation on the Code Summarization task.

% The fine-tuning and testing is done using the test split of each individual programming language.

% Both PLM will be trained with the mainstream dataset, Java and Python. The datasets and the two PLM are described in Section \ref{subsection:plm-dataset}.
% % while the two PLM are described in Section \ref{subsection:pre-trained models}. 
% For each downstream task, six programming languages (\textit{Python, Javascript, Ruby, Go, Java and PHP}) datasets are used to fine-tune the pre-trained PLM to create programming language specific downstream models. For example, the validation dataset of Ruby are used to fine-tune the code summarization task using the PLM pre-trained with Java and Python. After fine-tuning, a code summarization model will be created for generating comments for Ruby code. Other programming languages follow similarly. We note here that besides Java and Python, the other programming languages datasets are less popular, and Ruby is a low-resource programming language. To better understand the effects of PLM that are pre-trained with the mainstream programming languages, we also compare with PLM pre-trained with all the programming languages datasets.
\hfill 

\noindent \textbf{RQ2 Design:} Here, we compute the Performance-to-Time Ratio -- we measure the training time it takes to fine-tune a PLM and compare the time with its performance in the downstream tasks. 
For Code Summarization, the performance is measured in BLEU and METEOR, while for Code Search, the performance is measured in MRR.
The time and performance are all normalized within the range of 0 and 1 prior to computing the ratio. We normalized the training time by having the largest training time as the denominator of all training times --- this will have the effect of the largest training time having a normalized value of 1 and other training times scaled with respect to it. We normalized the performance in a similar manner, except that instead of the training time, we use the performance metric scores: BLEU, METEOR, and MRR.

% The pre-training of PLMs in RQ2 follows similarly to RQ1, except for the fine-tuning and testing processes, we use different programming languages.

% For RQ2, we also train RoBERTa and ELECTRA which will be used for the same downstream tasks described previously in RQ1. However, instead of using the mainstream datasets, Java and Python, to train the PLM, the programming language that will be used for fine-tuning the downstream task is used.
% % for each downstream task (which will be fine-tuned using other languages),  
% For example, the Ruby training dataset is used to pre-train a PLM and the pre-trained PLM will be used for fine-tuning the code summarization task, using the Ruby validation dataset.
\hfill 

\noindent \textbf{RQ3 Design:} We compare among the PLMs that were not pre-trained nor fine-tuned using the target language. We first pre-train the monolingual PLMs (except for Ruby). Then, we fine-tuned the monolingual PLMs using different monolingual datasets (except for Ruby). For each monolingual PLM, there are five fine-tuned models (based on the five monolingual datasets). For each monolingual PLM, we compare with all its fine-tuned models to find the best performing one. Then, among the best fine-tuned models in all the PLMs, we compare them to find the best model.

\hfill 

\noindent \textbf{RQ4 Design:} We compare the effects on the different code lengths on Code Summarization, similar to previous work \cite{lieditsum}. Additionally, we also compare the effects on the different code lengths on Code Search, which has not been studied. We segregate the target test data into four different code length groups, based on the length distribution of code fragments in the target programming language. The four groups are: 1) code length between 0 and first quartile, 2) code length between first and second quartile, 3) code length between second and third quartile, and 4) code length on and above third quartile. After segregating the test data into these four groups, we test all PLMs on each group. Specifically, for each PLM, we test all the fine-tuned models on each group. Among them, we compare to find the best fine-tuned model. We then compare the best fine-tuned models of each PLM to find the overall best model. Additionally, we also compare with the multilingual PLMs fine-tuned with the combined dataset which has reported having good performance \cite{ahmed2021multilingual}.  

\hfill 

\noindent \textbf{RQ5 Design} We propose a strategy (Section \ref{Section:Stratgey}) to select a set of programming languages that works well in multilingual PLMs. We fine-tuned the PLMs with this selected set of programming languages.

% For RQ3, it follows similar to both RQ1 and RQ2. However, there are some differences in the pre-training, fine-tuning and testing processes. For example, in the pre-training process, we use high-resource programming languages (Java and Python), and for fine-tuning and testing processes, we use low-resource programming languages.

% RQ3 can be viewed as a modification to the PLM in RQ1 and RQ2. Specifically, we create new PLM models based on the PLM trained in RQ1 and RQ2, by replacing the trained weight values with random values. In terms of fine-tuning, the same settings remain.

For all the RQs, we use CodeBERT and GraphCodeBERT, the two state-of-the-art PLMs for code, as our baseline. They are pre-trained and fine-tuned on the multilingual datasets.

\section{Selecting a Programming Language for Fine-tuning a PLM}
\label{Section:Stratgey}

% \begin{figure}
% \centering
%   \includegraphics[width=0.9\linewidth]{figures/Overview Select PL.png}
%   \caption{Overview of our proposed approach in selecting a compatible programming language that works well with the target PL.}
%   \label{fig:overview-select-pl} 
% \end{figure}

\subsection{Code Summarization}
\noindent\textbf{Overview} We consider a suitable programming language that can be used to fine-tune a PLM for a target programming language to have both similar semantics and textual properties compared to the target programming language. 
% Figure \ref{fig:overview-select-pl} presents an overview of our approach. 
We first train an embedding model using the whole multilingual dataset. Then, we compute semantic similarity between the individual monolingual datasets and the target programming language dataset. Afterward, we compute textual similarity between monolingual datasets and the target programming language dataset. Finally, we select the suitable programming language based on our proposed formula, which takes into account both the semantic and textual similarity scores.

\noindent\textbf{Semantic Similarity.} To detect similar semantics between programming languages, we train an embedding model (using the whole multilingual dataset) before computing the cosine similarity between the programming language and the target programming language. We made use of a recent paragraph embedding model that has reported having good performance in computing similarities between sentences by training the word embeddings jointly with bigram and trigram embeddings \cite{DBLP:conf/naacl/GuptaPJ19}. To train the embedding model, for every code function, we remove all line breaks and treat the code as a sequence of continuous word tokens --- this sequence of continuous word tokens can be viewed as a sequence of sentence. The trained embeddings model is then used to retrieve the embeddings of every monolingual code and the target programming language code for computing cosine similarity. Finally, between each programming language and the target programming language, we compute the average similarity score. We further normalized this score to be in the range of 0 and 1, by dividing all the values with the largest average similarity score.

\noindent\textbf{Textual Similarity} To detect similar text between a programming language and the target programming language, we made use of CCFinder \cite{kamiya2002ccfinder}, a token-based code clone detector to compute the textual similarity between the programming language and the target programming language. CCFinder is able to detect code clones in a variety of format, including C, C++, C\#, Cobol, Java, VB and plaintext. As our purpose is to detect “textual similarity” between cross programming languages code, we have used the plaintext mode in CCFinder.
Here, the textual similarity is the number of code clones detected. CCFinder transforms code fragments into suffix trees and uses them to detect exact and near-miss code clones. To have a more fine-grained detection, we set the minimum number of detected tokens to be 30, instead of the default 50. We perform code clone detection between every programming language and the target programming language. Finally, between each programming language and the target programming language, the number of code clones is computed. Similarly, we normalized this number to be in the range of 0 and 1, by dividing all the values with the largest number of code clones.

\noindent\textbf{Suitability Function} The suitability of a programming language is then formulated as:
\begin{align}
\small    \frac{Sim_{sem} + Sim_{text}}{2} \geq \theta 
\label{eqn:suitability}
\end{align}

where $Sim_{sem}$ refers to the normalized cosine similarity between the programming language and the target programming language, and $Sim_{text}$ refers to the normalized number of code clones between the programming language and the target programming language. Following previous work, we set $\theta$ to be 0.5 \cite{li2017cclearner}.

\subsection{Code Search}
Based on our empirical experiments, we have observed that PLMs fine-tuned with the combined multilingual dataset perform best in Code Search (Table \ref{tab:rq1-code-search}). Thus, for the Code Search task, we propose using the combined multilingual dataset to fine-tune the PLMs.

\section{Experimental Setup}
\label{Section:Experimental-Setup}
In this section, we describe in detail 1) the PLM and the dataset used for training the PLM, 2) the downstream tasks and the datasets used for fine-tuning the models for these downstream tasks, 3) the evaluation metrics for the downstream tasks, and 4) the qualitative analysis for Code Summarization in Section \ref{subsection:plm-dataset}, \ref{subsection:downstream-dataset}, \ref{subsection:evaluation-metrics}, and \ref{Subsection:human-comments-evaluation} respectively. 
% We note here that we fine-tune each downstream task using both PLM separately.

\subsection{The PLM and the Pretraining Dataset}
\label{subsection:plm-dataset}

% \textbf{Pre-trained Language Models}
% \label{subsection:pre-trained models}

% Explain RoBERTa and ELECTRA, and why we choose these two models, how they are trained, etc.

\textbf{RoBERTa} 
% (Robustly optimized BERT approach)
enhances the pre-training task of BERT \cite{bert} and achieved higher performance in many NLP tasks \cite{roberta}. 
It uses a bidirectional Masked Language Modeling (MLM) objective, where the model is trained to predict the masked tokens in the input text. 
In MLM, a small number of words are masked (15\%) and the model is trained to predict them \cite{clark2020electra}. 
We use RoBERTa as it is a common base model used in many PLM studies in SE such as CodeBERT and GraphCodeBERT \cite{feng2020codebert,guo2020graphcodebert}. We are unable to use CodeBERT or GraphCodeBERT for pre-training from scratch as their source code is close-sourced and not available.

% \textbf{PLM \#2: ELECTRA} pre-trains the text encoders as discriminators \cite{clark2020electra}. 
% ELECTRA introduces a new pre-training objective, known as Replaced Token Detection (RTD). 
% The RTD uses a discriminator to determine if a token in the input is \textit{real} or was it \textit{replaced} by another token. 
% We use ELECTRA in our study as it is reported to outperform many  state-of-the-art models in different NLP tasks. 
% % Additionally,
% The RTD pre-training task is also used in a previous study where the PLM is trained using code and has reported good performance \cite{feng2020codebert}.
% as well and therefore, we choose this model in our study. 

\begin{table}[htbp]
\caption{Dataset used for training PLM.}
\small
\begin{center}
\begin{tabular}{|c|c|c|}
\hline
\textbf{Language} & \textbf{bimodal DATA}& \textbf{unimodal CODES} \\
\hline
\textbf{Go} & 317,832 & 726,768 \\
\hline
\textbf{Java} & 500,754 & 1,569,889 \\
\hline
\textbf{Javascript} & 143,252 & 1,857,835 \\
\hline
\textbf{PHP} & 662,907 & 977,821  \\
\hline
\textbf{Python} & 458,219 & 1,156,085 \\
\hline
\textbf{Ruby} & 52,905 & 164,048	\\

\hline

\end{tabular}
\end{center}
\label{data:plm}
\end{table}

\noindent\textbf{Dataset.} As shown in Table \ref{data:plm}, we train RoBERTa using the CodeSearchNet data \cite{husain2019codesearchnet}, a dataset published by GitHub and Microsoft. 
It contains two different types of data: 1) parallel data of natural language-code pairs, known as bimodal data (column two) and 2) codes without paired natural language and natural language without paired codes, known as unimodal data (column three). 
% The former is referred to bimodata data and the latter unimodal data. 
Each unimodal code is a function without paired documentation.
% This dataset consists of over two million code and comment pairs, and over 4 million functions without documentation that is collected from open source libraries. 
The programming languages in this dataset are Go, Java, Javascript, PHP, Python and Ruby.
There are 2.1M bimodal data points and 6.4M unimodal codes.
This dataset is commonly used in previous studies \cite{feng2020codebert,husain2019codesearchnet,guo2020graphcodebert}.
% We use the code-comment pairs of the dataset to train the models (both PLM and the code summarization downstream task). 
% This is mainly related to the fact that code summarization requires both code and comment pairs. 
% The dataset is pre-processed by the publishers and the cleaning scripts are provided. 
% We use the same cleaned dataset in our experiments.
% CodeSearchNet is split into train, test, and validation sets, and we use the same splits to train our models. 
% Table \ref{data:codesearchnet} shows the statistics of the dataset.

\subsection{Downstream Tasks and Datasets}
\label{subsection:downstream-dataset}

% \begin{table}[htbp]
% \label{data:codesearchnet}
% \caption{Dataset for Comment Generation. The mainstream languages, Java and Python, are two of the major datasets. Ruby is a low resource language.}
% \begin{center}
% \begin{tabular}{|c|c|c|c|}
% \hline
% \textbf{Language} & \textbf{bimodal Data}& \textbf{unimodal Data} \\
% \hline
% \textbf{Go} & 317,832 & 14,242  \\
% \hline
% \textbf{Java} & 454,451 & 15,328 \\
% \hline
% \textbf{Javascript} & 123,889 & 8,253  \\
% \hline
% \textbf{PHP} & 523,712 & 26,015  \\
% \hline
% \textbf{Python} & 412,178 & 23,107  \\
% \hline
% \textbf{Ruby} & 48,791 & 2,209 	\\

% \hline

% \end{tabular}
% \label{tab:zero-shot}
% \end{center}
% \end{table}

\textbf{Task \#1: Code Summarization.}
The Code Summarization task is to generate textual summaries describing the code, where the input to the model is a code snippet and the output is a description of the code functionality in natural language. 
For fine-tuning, we followed the CodeBERT paper and used their published code -- for Code Summarization, an encoder-decoder framework is used to train a model to generate summaries, while for Code Search, the representation of [CLS] is used to measure the semantic relevance between the code and query \cite{feng2020codebert}. 
% It uses the CodeSearchNet \cite{husain2019codesearchnet} data for comment generation. 

\noindent\textbf{Dataset for Task \#1.} We leverage the same dataset as described in Section \ref{subsection:plm-dataset} but % CodeSearchNet data \cite{husain2019codesearchnet}, a dataset published by GitHub and Microsoft. 
% This dataset consists of over two million code and comment pairs, and over 4 million functions without documentation that is collected from open source libraries. 
% The programmings languages in this dataset are Python, Javascript, Ruby, Go, Java, and PHP.
% The dataset is commonly used in other studies including CodeBERT \cite{feng2020codebert}. 
use only the code-comment pairs to fine-tune the Code Summarization models.
% s (both PLM and the code summarization downstream task). 
% This is mainly related to the fact that code summarization requires both code and comment pairs. 
The dataset is pre-processed by the publishers and the cleaning scripts are provided. 
We use the same cleaned dataset in our experiments.
It is split into train, test, and validation sets, and we use the same split to train our models. 
Table \ref{data:codesearchnet} shows the statistics of the dataset. We note that 
% Java and Python are high-resource programming languages and 
Ruby is a low-resource programming language.

\begin{table}[htbp]

\caption{Dataset for Code Summarization.}
% Ruby is a low-resource language.}
\small
\begin{center}
\begin{tabular}{|c|c|c|c|}
\hline
\textbf{Language} & \textbf{Train}& \textbf{Valid}& \textbf{Test} \\
\hline
\textbf{Go} & 317,832 & 14,242 & 14,291 \\
\hline
\textbf{Java} & 454,451 & 15,328 & 26,909 \\
\hline
\textbf{Javascript} & 123,889 & 8,253 & 6,483 \\
\hline
\textbf{PHP} & 523,712 & 26,015 & 28,391 \\
\hline
\textbf{Python} & 412,178 & 23,107 & 22,176 \\
\hline
\textbf{Ruby} & 48,791 & 2,209 & 2,279	\\

\hline

\end{tabular}
\end{center}
\label{data:codesearchnet}
\end{table}

% \textcolor{red}{Not sure if code summarization is the right term. In the literature, there is a slight difference among the two tasks of code summarization and comment generation. Though, it seems there is no consensus on the term. }

\noindent\textbf{Task \#2: Code Search.}
The Code Search task is to find the most semantically related code from a collection of codes, given a natural language as the input.
% WBO detection is proposed by Pradel and Sen \cite{pradel2018deepbugs} as the task of finding the correctness of a binary operator in an expression. 
% WBO is used as a downstream task in previous PLM developed for code \cite{kanade2020learningCuBERT}. 
% For wrong binary operator,
For fine-tuning, we followed the CodeBERT paper \cite{feng2020codebert}. 

\noindent\textbf{Dataset for Task \#2.} 
We used a preprocessed version of the CodeSearchNet data where a correct pair of test data <docstring, code> is combined with a fixed set of 999 incorrect pairs of test data  \cite{feng2020codebert}. This data preprocessing is used for computing the Mean Reciprocal Rank (MRR). This dataset is commonly used in previous Code Search studies \cite{feng2020codebert,husain2019codesearchnet}.

% We use the ETH Py150 Open Corpus that is synthetically prepared by Kanade et al. \cite{kanade2020learningCuBERT}. 
% This dataset is synthetically prepared 
% from the ETH Py150 corpus \cite{raychev2014code} 
% It has been used for fine-tuning a PLM trained on code corpora for multiple source code related downstream tasks \cite{kanade2020learningCuBERT}. 
% For the WBO detection task, the data is prepared by replacing a random binary operator in each function with a different compatible binary operator. 
% Each function is included twice, a version containing the wrong binary operator, and
% another version containing the correct binary operator. 
% Additional information for each function is provided, including the label, where the function had originally been sourced from, and what operator had been replaced, if there is any. 
% Only the function and its corresponding label are used for the WBO detection task. 
% The dataset for the WBO detection task contains only Python functions.
% We use the same split created by Kanade et al.: 459,400 for training, 49,804 for validation, and 251,804 for test. 

\subsection{Evaluation Metrics}
\label{subsection:evaluation-metrics}

\textbf{The Code Summarization task} is evaluated using BLEU \cite{bleu} and METEOR \cite{meteor} where the generated summaries are compared against the ground truth comments. These metrics are commonly used in Code Summarization studies \cite{ase, rencos}. BLEU and METEOR scores are numbers between 0 and 1, and we report their percentages following previous work \cite{rencos, hgnn}. 

\noindent\textbf{BLEU} is a precision-based metric and measures the n-gram geometric precision 
% , $p_n$, 
between the generated summary and the ground truth summary \cite{papineni2002bleu}. 
% A low BLEU score indicates that the generated summary does not closely resemble the human-written ground-truth summary. 
% The brevity penalty, $BP$, penalizes the shorter candidate sentences of length $c_l$.  $r_l$ is the length of the reference text. 

% \begin{equation}
%   BLEU =BP.exp \sum_{N=1}^{4} \frac{1}{N} \log p_n ,\hspace{4pt}
%   BP=
%     \begin{cases}
%       1 & \text{if}\ c_l>r_l \\
%       e^{\frac{1 - r_l}{c_l}}, & \text{if}\ c_l \leqslant r_l\\\label{brevity}
%     \end{cases}
% \end{equation}

% \noindent\textbf{ROUGE-L} is a recall-based metric that computes the longest common subsequent between the generated summary and the ground truth summary \cite{lin2004rouge}.
% and is calculated as follows. 
% X and Y are the sentences with length $m$ and $n$, and ROUGE-L computes the F-score of the longest common subsequent (LCS) among them, where $R_{LCS} = \frac{LCS(X,Y)}{m}$, $P_{LCS} = \frac{LCS(X,Y)}{n}$, and $\beta =\frac{P_{LCS}}{R_{LCS}}$. 

% \begin{equation}
% F_{LCS} = \frac{(1+\beta^{2})R_{LCS}P_{LCS}}{R_{LCS} + \beta^{2}P_{LCS}}\label{rouge}
% \end{equation}

%\textcolor{red}{METEOR was designed to overcome the problem of lack of recall, explicit word matching of the BLEU score} %\cite{meteor}. 
\noindent\textbf{METEOR} 
% calculates the score using an alignment approach 
measures the alignment between the generated summary and the ground-truth summary using exact, stem, and synonym matches between words and phrases \cite{banerjee2005meteor}.
% using the following formula: 
% \begin{equation}
% METEOR = (1 -\gamma.frag^{\beta}). \frac{P.R}{\alpha.P + (1-\alpha).R}\label{meteor}
% \end{equation}

% where $P$ and $R$ are the precision and recall values, and $\alpha$, $\beta$ and $\gamma$ are set to their default values $0.9$, $3.0$ and $0.5$, respectively.

% \noindent\textbf{CIDEr} evaluates the correctness of the generated summary \cite{rencos, cocogum} using the Term Frequency-Inverse Document Frequency (TF-IDF) weighting of the tokens and cosine similarity of the generated summary and ground truth summary \cite{vedantam2015cider}.
% , as shown in Equation \ref{cider-eqn-1} and \ref{cider-eqn-2}. $N$ is set to 4 and $g^{n}(c_{i})$ is the TF-IDF weighted vector. 

% \begin{equation}
% CIDEr_n(C_i, S_i) = \frac{1}{m}\sum_{j} \frac{g^{n}(c_{i}).g^{n}(s_{ij})}{\left|\left|g^{n}(c_{i})\right|\right|\left|\left|g^{n}(c_{i})\right|\right|}
% \label{cider-eqn-1}
% \end{equation}

%  and 

% \begin{equation}
% % \label{cidereqn2}
%  CIDEr(C_i, S_i) = \sum_{1}^{N}\frac{1}{N}CIDEr_n(C_i, S_i)
% \label{cider-eqn-2}
% \end{equation}

% CIDEr has positive real value and can be above 1.
% % We report CIDEr as is. 
% The higher the score, the better the generated results.  

\noindent\textbf{The Code Search task} is a search task and it is evaluated using the Mean Reciprocal Rank (MRR). 

\noindent\textbf{MRR} evaluates the Code Search task where it produces a list of possible responses to the code query. The reciprocal rank of the code query response is the multiplicative inverse of the rank of the first correct answer, and the MRR is the average of the reciprocal ranks of results for the code queries \cite{voorhees1999trec,radev2002evaluating}.

\subsection{Qualitative Analysis for Code Summarization}
\label{Subsection:human-comments-evaluation}
					
For Code Summarization, we further conducted a qualitative analysis. We randomly select 800 generated summaries (for each PLM) along with their original code, 100 pairs for each of the best performing monolingual and multilingual PLMs, following prior research~\cite{liu2019text, grusky2018newsroom}. Amazon Mechanical Turk (MTurk) workers were hired to rate the quality of the generated summaries. The MTurkers rated the summary voluntarily, and for each rated summary, the MTurkers are given a compensation of one cent. There are three different annotators for each generated code summary, and different generated code summaries may not get the same annotators, due to the randomness of the MTurkers' assignments. We also ask the MTurkers if they understand the code and we only accept those ratings when the MTurkers have stated that they do understand the code. This will ensure that the annotations are proper and they will not be biased towards specified annotators. In total, 2,088 MTurkers have participated in the study.  We used four common criteria to evaluate the summarization quality \cite{liu2019text}: \\
\textbf{Informativeness} How well the summary capture the key points of the code? \\
\textbf{Relevance} Are the summary details consistent with in the code? \\
\textbf{Fluency} Are the summaries well-written and grammatically correct? \\ 
\textbf{Comprehension} Can the summaries help in understanding the code? \\ 
Three different workers were required to rate each summary between one and five, where one is the worst and five is the best. 

\section{Results}
\label{Section:Results}
In this section, we will first discuss the result of Code Summarization before Code Search.

\subsection{RQ1:  Does training and fine-tuning on the individual programming languages help to improve the performance over multilingual PLMs that are fine-tuned on multilingual datasets?}

\subsubsection{Code Summarization}

\begin{table*}[htbp]
\caption{Code Summarization using different PLMs (trained using monolingual dataset) fine-tuned on the monolingual and combined multilingual datasets for Ruby. Here, we only show the fine-tuned model of each PLM having the best BLEU scores e.g., PLM$_{Ruby}$/Combined refers to the PLM pre-trained using the Ruby dataset and fine-tuned on the combined multilingual dataset. The left value in the bracket shows the average score while the right value shows the standard deviation. 
% We also compare with the multilingual PLMs fine-tuned on the combined multilingual dataset.
}
% \small
\begin{center}
\begin{tabular}{cccccccc}
\hline
\textbf{PLM} & \textbf{BLEU-1} & \textbf{BLEU-2} & \textbf{BLEU-3} & \textbf{BLEU-4} & \textbf{METEOR}  \\
\hline
PLM$_{Ruby}$/Combined &  16.5 (15.0/2.3) & 7.9 (6.3/1.4) & 4.3 (3.1/0.9)  & 2.6 (1.7/0.6)  & 12.5 (11.0/0.1)   \\
PLM$_{Javascript}$/Combined & 15.7 (14.6/1.7)  & 7.4 (6.2/1.5) & 4.0 (3.2/1.1) & 2.4 (1.8/0.1) & 12.4 (11.2/1.1) 	\\
PLM$_{PHP}$/Combined & 16.1 (14.2/2.7) & 7.9 (6.1/1.6)  & 4.5 (3.1/1.0) & 2.8 (1.8/0.7) & 12.8 (11.5/0.9) \\
PLM$_{Java}$/Combined & 16.1 (14.1/2.4) & 8.0 (6.3/1.6) & 4.6 (3.3/1.1) & 2.8 (1.9/0.8) & 12.7 (11.5/1.0)  \\
PLM$_{Python}$/Combined & 15.7 (15.1/2.2) & 7.5 (6.7/1.3) & 4.2 (3.6/0.9) & 2.6 (2.0/0.6) & 12.5 (11.8/0.7)  \\
PLM$_{Go}$/Combined & 14.7 (15.0/2.7) & 7.4 (6.4/1.6) & 4.2 (3.3/1.1) & 2.6 (1.8/0.7) & 12.6 (11.3/1.0)  \\ \hline
CodeBERT/Combined & 15.4 & 8.2  & 4.9 & 3.1  & 13.2   \\
GraphCodeBERT/Combined & 16.5 & 9.1 & 5.8  & 3.9  & 13.5  \\
\hline

\end{tabular}
\label{tab:rq1-code-summarization}
\end{center}
\end{table*}

Table \ref{tab:rq1-code-summarization} shows the best performing monolingual PLM. The first column shows the PLM and the second to sixth column show the BLEU and the METEOR metrics. 

We observe that for all the PLMs, fine-tuning on the combined dataset gives the best performance. Comparing the monolingual and multilingual PLMs i.e., CodeBERT and GraphCodeBERT, the latter has better performance.

% Table \ref{table:qualitative} shows the survey results from the Amazon MTurkers on the generated summaries given the code. The majority of the MTurkers have (\%) Java experience and (\%) Python experience between 1 to 5 years, in the Java survey and Python survey, respectively. 

\begin{table}[t!]
% 	\small
\caption{Qualitative results from the MTurker studies.}
\small
	\centering
	\begin{tabular}{cccccc}
	
	\hline
	
    \multirow{1}{*}{\textbf{}} &
      \multicolumn{1}{c}{\textbf{Info.}} &
      \multicolumn{1}{c}{\textbf{Rel.}} &
      \multicolumn{1}{c}{\textbf{Flu.}} &
      \multicolumn{1}{c}{\textbf{Compre.}} 
      \\
      \hline

    PLM$_{Ruby}$/Combined & 4.40 & 4.57 & 4.63 & 4.62 \\ 
    PLM$_{Javascript}$/Combined & 4.46  & 4.47 & 4.54 & 4.57 \\ 
    PLM$_{PHP}$/Combined & 4.49  & 4.45 &  4.53 & 4.58 \\ 
    PLM$_{Java}$/Combined & 4.42 & 4.40 & 4.44 & 4.45 \\ 
    PLM$_{Python}$/Combined & 4.47 & 4.42 & 4.53 & 4.63\\ 
    PLM$_{Go}$/Combined & 4.45  & 4.45 & 4.42  & 4.49 \\ \hline
    CodeBERT/Combined & 4.41 & 4.46 & 4.46  & 4.41 \\ 
    GraphCodeBERT/Combined & 4.33  & 4.35 & 4.44  & 4.39 \\ 
    \hline   
    
	\end{tabular}
	\centering
	\label{tab:rq1-code-summarization-mturks}
\end{table}

Table \ref{tab:rq1-code-summarization-mturks} shows the annotation results from the Amazon MTurkers. Generally, the MTurkers find that the generated summaries of all the PLMs are informative, relevant, fluent, and that they can help in the understanding of the code -- the summaries are rated above 4 in these areas. 
% The MTurkers find that the generated summaries from PLM$_{PHP}$/Combined, PLM$_{Ruby}$/Combined, and PLM$_{Ruby}$/Combined, are most informative, relevant and fluent respectively, and that the generated summaries from PLM$_{Python}$/Combined can most help them to understand the code.

\subsubsection{Code Search}

\begin{table}[htbp]
\caption{Code Search using monolingual PLMs fine-tuned on the monolingual and combined multilingual dataset. Here, we only show the best performing fine-tuned model of each PLM. The left value in the bracket shows the average score while the right value shows the standard deviation. 
% We also compare with the multilingual PLMs fine-tuned on the combined multilingual dataset.
% The first column shows the PLM pre-trained on the individual programming language while the third and fourth columns show the percentage improvement over CodeBERT and GraphCodeBERT respectively, when the PLM is fine-tuned on the combined multilingual dataset. Surprisingly, we observed that all the PLMs pre-trained on the individual programming language outperformed the multilingual CodeBERT in MRR between 5.02\% and 35.1\%, and the multilingual GraphCodeBERT in MRR between \% and \%.
}
\small
\begin{center}
\begin{tabular}{ccccc}
\hline
\textbf{} & \textbf{} & \textbf{Improve} & \textbf{Improve} \\
\textbf{PLM} & \textbf{MRR} & \textbf{CodeBERT} & \textbf{GCodeBERT} \\
\hline
PLM$_{Ruby}$/Combined & 0.57(0.41/0.12) & +35.1\% & +32.6\%\\
PLM$_{Javascript}$/Combined & 0.44(0.32/0.07) & +5\% & +2.3\%	\\
PLM$_{PHP}$/Combined & 0.47(0.29/0.09) & +9.9\% & +9.3\% \\
PLM$_{Java}$/Combined & 0.46(0.29/0.10) & +9.8\% & +7\%\\
PLM$_{Python}$/Combined & 0.48(0.31/0.10) & +12.7\% & +11.6\%\\
PLM$_{Go}$/Combined & 0.46(0.24/0.13) & +9.8\% & +7\%\\
% CodeBERT & 0.42 & - & -\\
% GraphCodeBERT & 0.43 & - & -\\
\hline

\end{tabular}
\label{tab:rq1-code-search}
\end{center}
\end{table}

Table \ref{tab:rq1-code-search} shows the Code Search results of the best performing fine-tuned model of each PLM. The first and second column show the model and its MRR scores. The third and fourth column show the improvement over CodeBERT and GraphCodeBERT respectively. Similar to the Code Summarization task, we observe that for all the PLMs, fine-tuning on the combined multilingual dataset give the best performance. However, we were surprised to observe that all the monolingual PLMs 
outperformed CodeBERT in MRR between 5\% and 35.1\%, and GraphCodeBERT in MRR between 2.3\% and 32.6\%.

We conducted a Mann Whitney U-test between using the monolingual datasets and the multilingual datasets for fine-tuning, and the p-value is 0.00256 $<$ 0.05, showing that our experiments are statistically significant. 

\begin{tcolorbox}
Mutilingual PLMs fine-tuned on the combined multilingual dataset perform better than monolingual PLMs in Code Summarization whereas monolingual PLMs perform better than multilingual PLMs on Code Search.
\end{tcolorbox}

\subsection{RQ2: Which PLM has the best Performance-to-Time Ratio?}

\begin{figure}
\centering
   \includegraphics[width=0.8\linewidth]{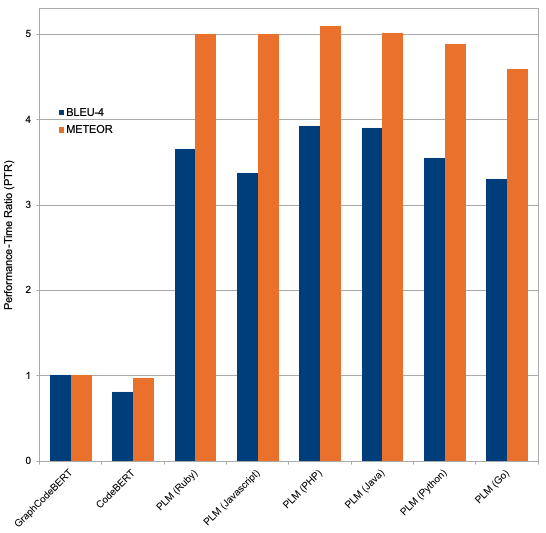}
   \caption{Performance-to-Time Ratio (PTR) for Code Summarization. Monolingual PLMs have higher BLEU-4 and METEOR PTR.}
   \label{fig:ptr-code-summarization} 
\end{figure}

\begin{figure}
\centering
   \includegraphics[width=0.8\linewidth]{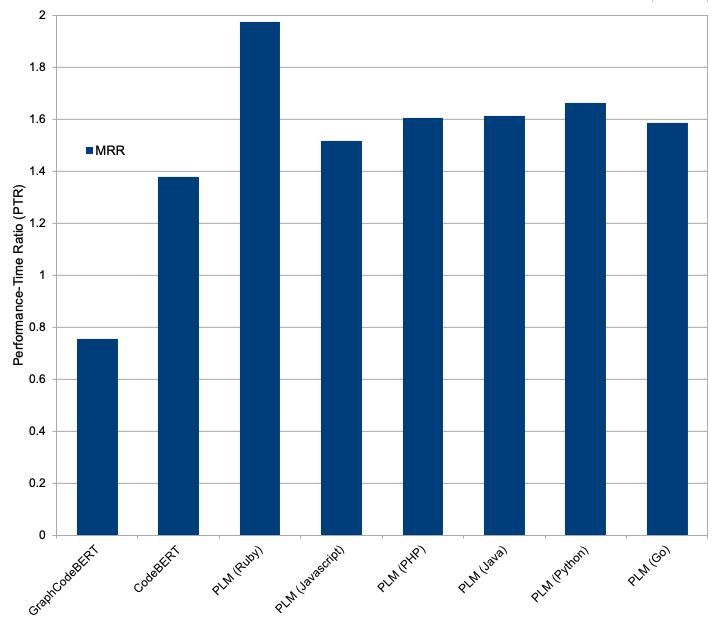}
   \caption{PTR for Code Search. The monolingual PLMs have shown a higher MMR Performance-to-Time Ratio (PTR).}
   \label{fig:ptr-code-search}
\end{figure}

% \begin{figure}
% \centering
% \begin{subfigure}[b]{0.45\textwidth}
%   \includegraphics[width=1\linewidth]{figures/PTR (BLEU-4 and METEOR).png}
%   \caption{PTR for Code Summarization}
%   \label{fig:ptr-code-summarization} 
% \end{subfigure}

% \begin{subfigure}[b]{0.5\textwidth}
%   \includegraphics[width=1\linewidth]{figures/PTR (MRR).png}
%   \caption{PTR for Code Search}
%   \label{fig:ptr-code-search}
% \end{subfigure}

% \caption{(a) For Code Summarization, the individual programming language PLMs have shown a higher BLEU-4 and METEOR Performance-to-Time Ratio (PTR) than CodeBERT. For CodeBERT, it takes between 4.91 to 5.44 times of additional hours to fine-tune (b) For Code Search, the individual programming language PLMs have shown a higher MMR Performance-to-Time Ratio (PTR) than CodeBERT.}
% \end{figure}

\subsubsection{Code Summarization}
Figure \ref{fig:ptr-code-summarization} shows the Performance-to-Time Ratio of the PLMs: the ratio of BLEU-4 and the model fine-tuning time, and the ratio of METEOR and the model fine-tuning time. 
% Note that all the BLEU-4, METEOR and the fine-tuning time are normalized to a value between 0 and 1 before the ratios are computed. To normalize the BLEU-4 values, we divide all the BLEU-4 values with the largest BLEU-4 value. The normalization for METEOR is computed similarly.
% GraphCodeBERT and CodeBERT refer to the multilingual PLMs while the other PLMs refer to the monolingual PLMs. 
We observe that for the multilingual PLMs, they have lower Performance-to-Time Ratio than the other PLMs. This shows that although the multilingual PLMs have higher performance, it would take much longer to fine-tune them.
For the monolingual PLMs, we observe that many of them have much larger Performance-to-Time Ratio than the multilingual PLMs, and monolingual PLM trained on the PHP dataset has the highest scores. 

\subsubsection{Code Search}
Figure \ref{fig:ptr-code-search} shows the Performance-to-Time Ratio of MRR and the model fine-tuning time. Similarly, the MRR and the model fine-tuning time are normalized to a value between 0 and 1 before the ratios are computed.
GraphCodeBERT and CodeBERT refer to the multilingual PLMs while the other PLMs refer to the monolingual PLMs. Similar to Code Summarization, we observe that the monolingual PLMs have higher Performance-to-Time  Ratio than the multilingual PLMs. For the monolingual PLMs, we observed that the monolingual PLM trained on the Ruby dataset has the highest score.

We conducted a Mann Whitney U-test between using the monolingual datasets and the multilingual datasets for the Performance-to-Time Ratio, and the p-value is 0.00256 $<$ 0.05, showing that our experiments are statistically significant. 

\begin{tcolorbox}
Monolingual PLMs fine-tuned on the combined dataset have the best Performance-to-Time Ratio. Researchers should consider choosing monolingual PLMs if they were to pre-train a PLM from scratch.
\end{tcolorbox}

\subsection{RQ3: What are the best settings for zero-shot downstream tasks? }

\subsubsection{Code Summarization}

\begin{table*}[htbp]
\caption{Zero-shot Code Summarization using different monolingual PLMs (other than Ruby) fine-tuned on the monolingual dataset (other than Ruby). Here, we only show the best performing fine-tuned model of each PLM (the average and standard deviation scores of other fine-tuned models are shown as the left and right values within the braces) e.g., PLM$_{Javascript}$/Python refers to the PLM pre-trained using the Javascript dataset and fine-tuned on the Python dataset.}
% \small
\begin{center}
\begin{tabular}{cccccccc}
\hline
\textbf{PLM} & \textbf{BLEU-1} & \textbf{BLEU-2} & \textbf{BLEU-3} & \textbf{BLEU-4} & \textbf{METEOR}  \\
\hline
PLM$_{Javascript}$/Python & 14.8 (13.8/0.9) & 7.3 (5.7/1.3) & 4.0 (2.8/1.0) & 2.4 (1.5/0.7)  & 12.3 (10.9/1.2/)  	\\
PLM$_{PHP}$/Python & 16.2 (13.5/3.0) & 7.6 (5.6/1.6) & 4.0 (2.8/0.9) & 2.4 (1.5/0.6) & 12.5 (11.3/0.8) \\
PLM$_{Java}$/Python & 16.2 (13.5/2.4) & 7.9 (5.7/1.5) & 4.5 (2.9/1.0)  & 2.8 (1.7/0.7)  & 12.6 (11.2/0.1)   \\
PLM$_{Python}$/Python & 15.7 (14.3/1.9) & 7.6 (6.2/1.2) & 4.2 (3.2/0.8) & 2.6 (1.8/0.5) & 12.5 (11.6/0.8)  \\
PLM$_{Go}$/Python & 16.4 (14.3/2.5) & 7.9 (5.9/1.5) & 4.3 (2.9/1.0) & 2.5 (1.5/0.7) &  12.4 (11.0/0.9) \\
\hline

\end{tabular}
\label{tab:rq3-code-summarization}
\end{center}
\end{table*}

Table \ref{tab:rq3-code-summarization} shows the zero-shot Code Summarization using the monolingual PLMs. The first column shows the PLM while column two to six show the BLEU and METEOR scores respectively. For all the PLMs, we observe that PLMs that are fine-tuned on the Python dataset has the best performance over other monolingual datasets. The average and standard deviation of the other fine-tuned models for each PLM is shown on the left and right values inside the braces.

\subsubsection{Code Search}

\begin{table}[htbp]
\caption{Zero-Shot Code Search using different monolingual PLMs (other than Ruby) fine-tuned on the monolingual dataset (other than Ruby). Here, we only show the best performing fine-tuned model of every PLM.}
\small
\begin{center}
\begin{tabular}{ccccc}
\hline
\textbf{PLM} & \textbf{MRR} \\
\hline
PLM$_{Javascript}$/Python & 0.37 (0.29/0.05)	\\
PLM$_{PHP}$/Python & 0.31 (0.25/0.05) \\
PLM$_{Java}$/Python & 0.34 (0.24/0.07) \\
PLM$_{Python}$/Python & 0.36 (0.26/0.08) \\
PLM$_{Go}$/Python & 0.36 (0.19/0.11) \\
\hline

\end{tabular}
\label{tab:rq3-code-search}
\end{center}
\end{table}

Table \ref{tab:rq3-code-search} shows the zero-shot results for Code Search using the monolingual PLMs. The first column shows the PLM while the second column shows the MRR scores. Interestingly, we observed that similar to the zero-shot settings in Code Summarization, fine-tuning on the Python dataset has the best MRR performance. 

We conducted a Mann Whitney U-test between using the datasets containing Ruby and the datasets excluding Ruby, and the p-value is 0.00604 $<$ 0.05, showing that our experiments are statistically significant.

\begin{tcolorbox}
For the zero-shot settings, we observed that PLMs fine-tuned on the Python dataset has the best performance. Researchers should consider using the Python dataset to fine-tune PLMs for Ruby.
\end{tcolorbox}

\subsection{RQ4: What effect does the PLMs have on different code length?}

\subsubsection{Code Summarization}

\begin{table*}[htb!]
\caption{Effects of PLMs on different code lengths fine-tuned on monolingual and combined multilingual dataset. Here, for each monolingual PLM, we show only the fine-tuned model of each PLM having the best BLEU scores e.g., PLM$_{Javascript}$/Combined refers to the PLM pre-trained using the Javascript dataset and fine-tuned on the combined multilingual dataset.}
\small
\begin{center}
\begin{tabular}{cccccccc}

\hline
\textbf{PLM} & \textbf{BLEU-1} & \textbf{BLEU-2} & \textbf{BLEU-3} & \textbf{BLEU-4} & \textbf{METEOR}  \\

\hline
\multicolumn{6}{c}{\textbf{0 - 1st Quartile Code Length}}\\
\hline

%\hline
PLM$_{Ruby}$/Combined & 18.2  & 8.5  & 4.7  & 3.0  & 13.2 \\
PLM$_{Javascript}$/Python & 16.7  & 8.2  & 4.6  & 2.9 & 12.9 	\\
PLM$_{PHP}$/Combined & 18.3  & 9.1  & 5.4 & 3.6 & 13.8  \\
PLM$_{Java}$/Combined & 18.8  & 9.2 & 5.2 & 3.3  & 13.7   \\
PLM$_{Python}$/Ruby & 21.0  & 10.2  & 5.8  & 3.3  & 13.0   \\
PLM$_{Go}$/Combined & 17.7  & 8.8  & 5.1  & 3.4  & 13.7   \\ \hline
CodeBERT/Combined & 17.8  & 9.4  & 5.5  & 3.4  & 14.0    \\
GraphCodeBERT/Combined & 17.8  & 9.4  & 5.5  & 3.4  & 14.0   \\
\hline

% \hline
\multicolumn{6}{c}{\textbf{1st - 2nd Percentile Code Length}}\\
\hline

%\hline
PLM$_{Ruby}$/Python & 16.4  & 7.9 & 4.3  & 2.6  & 12.7\\
PLM$_{Javascript}$/Ruby & 18.5  & 8.9  & 5.2 & 3.3 & 12.2 	\\
PLM$_{PHP}$/Combined & 17.2  & 8.5  & 4.9 & 3.1 & 13.2  \\
PLM$_{Java}$/Combined & 18.0 & 9.3 & 5.5 & 3.5  & 13.2   \\
PLM$_{Python}$/Python & 17.4  & 8.7  & 5.2  & 3.4  & 13.2    \\
PLM$_{Go}$/Ruby & 19.8  & 9.1  & 5.1  & 2.9  & 11.7  \\ \hline
CodeBERT/Combined & 16.6  & 9.0  & 5.6  & 3.7  & 13.9    \\
GraphCodeBERT/Combined & 17.2  & 9.4  & 6.0  & 4.1  & 14.0   \\
\hline

% \hline
\multicolumn{6}{c}{\textbf{2nd - 3rd Quartile Code Length}}\\
\hline

%\hline
PLM$_{Ruby}$/Combined & 14.2  & 7.0 & 4.1  & 2.6  & 11.7 \\
PLM$_{Javascript}$/Ruby & 15.4  & 6.9  & 3.9 & 2.3 & 10.1 	\\
PLM$_{PHP}$/Python & 14.1  & 7.0  & 3.9 & 2.4 & 12.0  \\
PLM$_{Java}$/Python & 13.5 & 6.9 & 4.2 & 2.7  & 12.0   \\
PLM$_{Python}$/Ruby & 16.3  & 7.7  & 4.3  & 2.5  & 11.5   \\
PLM$_{Go}$/Python & 14.2  & 7.0  & 3.9  & 2.3  & 11.8  \\ \hline
CodeBERT/Combined & 16.4  & 7.0  & 3.7  & 2.0 & 11.2   \\
GraphCodeBERT/Combined & 13.7  & 7.7  & 4.9  & 3.4  & 12.9   \\
\hline

% \hline
\multicolumn{6}{c}{\textbf{3rd Quartile - Max Code Length}}\\
\hline

%\hline
PLM$_{Ruby}$/Combined & 16.2  & 7.5 & 4.1  & 2.2  & 11.9 \\
PLM$_{Javascript}$/Python & 14.4  & 6.7  & 3.5 & 2.0 & 11.7 	\\
PLM$_{PHP}$/Combined & 15.1  & 7.1  & 3.9 & 2.2 & 11.9  \\
PLM$_{Java}$/Combined & 14.5 & 6.7 & 3.5 & 1.9  & 11.7   \\
PLM$_{Python}$/Combined & 14.6  & 6.9  & 4.0  & 2.5  & 11.6   \\
PLM$_{G}$/Python & 15.1  & 7.3  & 4.2  & 2.5  & 11.4  \\ \hline
CodeBERT/Combined & 14.3  & 7.4  & 4.3  & 2.7 & 12.2   \\
GraphCodeBERT/Combined & 14.7  & 7.8  & 4.9  & 3.2  & 12.3   \\
\hline

\end{tabular}
\label{tab:rq4-code-summarization}
\end{center}
\end{table*}

Table \ref{tab:rq4-code-summarization} shows the effects of the different code lengths when the PLMs are used on Code Summarization. For code lengths within the first percentile and code lengths above the third percentile, we observe that majority of the PLMs have the best performance when fine-tuned on the combined multilingual datasets. However, we oberved that for the other code lengths, some of the PLMs perform better when fine-tuned on Ruby or Python dataset.  

\subsubsection{Code Search}

Table \ref{tab:rq4-code-search} shows the effects of the different code lengths when the PLMs are used in Code Search. We observed that for all the different PLMs, fine-tuning on the combined multilingual dataset gave the best performance in all code lengths. Among the PLMs, there is very little variation in the MRR scores on the different code lengths i.e., their scores differ within +/-0.05 and majority of them are within +/-0.03.

\begin{tcolorbox}
For different code lengths, we observed similar performance when the PLMs are tested on all the code lengths.
\end{tcolorbox}

\subsection{RQ5: How effective is our strategy to decide in advance a language that can work well for a target low-resource language?}
\label{subsection:results-strategy-effectiveness}

\subsubsection{Code Summarization}

Based on the suitability equation \ref{eqn:suitability}, Python, PHP, Java, Javascript and Go have scores of 0.55, 0.92, 0.14, 0.52 and 0.46, when compared to Ruby, respectively. Python, PHP and Go are over 0.5 and thus we selected them for fine-tuning. Ruby is also included as it is the target language. Table \ref{tab:rq5-code-summarization} shows the performance of BLEU and METEOR when the PLMs are fine-tuned on the proposed set of programming languages. We observe improvement in BLEU and METEOR on both monolingual and multilingual PLMs. For BLEU-1, BLEU-2, BLEU-3, BLEU-4 and METEOR, the improvement ranges from 0.6\% to 4.3\%, 1.1\% to 13.5\%, 11.9\% to 30\%, 3.8\% to 41.7\%, and 1.5\% to 5.6\%, respectively.

% \begin{figure}
% \centering
%   \includegraphics[width=1\linewidth]{figures/Code Length - BLEU-4 and METEOR.pdf}
%   \caption{Effects of BLEU and METEOR on different code length for Code Summarization. Our proposed approach to select a dataset for fine-tuning has improved performance in both BLEU and METEOR over multilingual CodeBERT on different code length.}
%   \label{fig:code-length-code-summarization} 

% \end{figure}

\subsubsection{Code Search}

From Table \ref{tab:rq1-code-search}, we observed that the PLMs fine-tuned on the combined multilingual dataset has the best performance in MRR. Specifically, we observed that all the monolingual PLMs have better performance in MRR than the multilingual PLMs. The improvement over CodeBERT and GraphCodeBERT ranges from 5\% to 35.1\% and 2.3\% to 32.6\%, respectively.

\begin{tcolorbox}
Our proposed strategies in selecting PLs for fine-tuning downstream tasks are effective. For Code Summarization, the improvement ranges from 0.6\% to 41.7\% in BLEU, and from 1.5\% to 5.6\% in METEOR. For Code Search, the MRR scores improve from 2.3\% to 35.1\%. 
\end{tcolorbox}

\section{Discussions}
\label{Section:Discussion}

\noindent\textbf{Which PLM should I use?} In this work, we studied the performance impact of monolingual and multilingual PLMs. We found that although multilingual PLMs show good performance in the automatic metrics, the MTurkers do not find any major difference between the summaries generated from monolingual and multilingual PLMs for Code Summarization in the qualitative study. Furthermore, the Performance-to-Time Ratio of the PLMs also suggest that monolingual PLMs are more efficient. In the case of Code Search, we observed that monolingual PLMs have better performance than multilingual PLMs. Thus, considering both efficiency and performance, we propose that developers do not merely use the multilingual PLMs in their task, but to also compare with the monolingual PLMs. 

\noindent\textbf{Different strategy in selecting a suitable PL for different do-wnstream tasks}
In our experiment, we have observed that the strategies in selecting a suitable programming language for Code Summarization and Code Search works well in their respective tasks. However, we also observed that the strategy for Code Summarization may not work as well in Code Search, and vice-versa. We believe that depending on the task, a different strategy may be required. Nonetheless, we showed that our proposed strategies (Section \ref{Section:Stratgey}) are effective and we still recommend developers to adopt our proposed strategies in any of their tasks before attempting to come out with a new one.

\noindent\textbf{Non-exact Code Duplicates in CodeSearchNet Dataset} We note that non-exact code duplicates (Type 2-4 code clones) may exist in the CodeSearchNet dataset. We believe that excluding them during pre-training and fine-tuning may not necessarily yield more robust models, and that the PLMs can benefit from learning more diverse code structure using the dataset that contains non-exact code duplicates.

\begin{table*}
\caption{Effects of PLMs on different code lengths fine-tuned on the monolingual dataset. Here, we only show the best performing fine-tuned model of each PLM e.g., PLM$_{Javascript}$/Python refers to the PLM pre-trained using the Javascript dataset and fine-tuned on the Python dataset.}

% \small
\begin{center}
\begin{tabular}{cc|cc|cc|cc}
\hline

\multicolumn{2}{c|}{0 - 1st Quartile} & 
\multicolumn{2}{c|}{1st - 2nd Quartile} &
\multicolumn{2}{c|}{2nd - 3rd Quartile} &
\multicolumn{2}{c}{3rd Quartile - Max} \\

\textbf{PLM} & \textbf{MRR} & \textbf{PLM} & \textbf{MRR} & \textbf{PLM} & \textbf{MRR} & \textbf{PLM} & \textbf{MRR} \\
\hline

\hline
PLM$_{Ruby}$/Comb. & 0.57 & PLM$_{Ruby}$/Comb. & 0.59 & PLM$_{Ruby}$/Comb. & 0.58 & PLM$_{Ruby}$/Comb. & 0.55 \\
PLM$_{Javascript}$/Comb. & 0.44 & PLM$_{Javascript}$/Comb. & 0.47 & PLM$_{Javascript}$/Comb. & 0.44 & PLM$_{Javascript}$/Comb. & 0.43	\\
PLM$_{PHP}$/Comb. & 0.47 & PLM$_{PHP}$/Comb. & 0.47 & PLM$_{PHP}$/Comb. & 0.48 & PLM$_{PHP}$/Comb. & 0.44 \\
PLM$_{Java}$/Comb. & 0.35 & PLM$_{Java}$/Comb. & 0.34 & PLM$_{Java}$/Comb. & 0.32 & PLM$_{Java}$/Comb. & 0.33 \\
PLM$_{Python}$/Comb. & 0.48 & PLM$_{Python}$/Comb. & 0.48 & PLM$_{Python}$/Comb. & 0.47 & PLM$_{Python}$/Comb. & 0.48\\
PLM$_{Go}$/Comb. & 0.47 & PLM$_{Go}$/Comb. & 0.48 & PLM$_{Go}$/Comb. & 0.45 & PLM$_{Go}$/Comb. & 0.43 \\ \hline
CodeBERT/Comb. & 0.44 & CodeBERT/Comb. & 0.42 & CodeBERT/Comb. & 0.42 & CodeBERT/Comb. & 0.42\\
GCodeBERT/Comb. & 0.43 & GCodeBERT/Comb. & 0.44 & GCodeBERT/Comb. & 0.46 & GCodeBERT/Comb. & 0.41\\
\hline

\end{tabular}
\label{tab:rq4-code-search}
\end{center}
\end{table*}

\begin{table*}
\caption{Code Summarization using different PLMs fine-tuned on a subset of selected languages (\textit{Ruby + Python + PHP + Go}) and tested on Ruby. 
% PLM$_{Ruby}$ refers to the PLM pre-trained on Ruby. 
The value inside the braces symbolized the percentage improvement over PLMs fine-tuned on the combined multilingual dataset. There is an improvement in all the metrics among the different PLMs.}
% \small
\begin{center}
\begin{tabular}{cccccccc}
\hline
\textbf{PLM} & \textbf{BLEU-1} & \textbf{BLEU-2} & \textbf{BLEU-3} & \textbf{BLEU-4} & \textbf{METEOR}  \\
\hline
PLM$_{Ruby}$ &  15.7 $(-4.8\%)$ & 7.7 $(-2.5\%)$ & 4.3 $(+0\%)$ & 2.7 $(+3.8\%)$ & 12.7 $(+2.7\%)$  \\
PLM$_{Javascript}$ & 15.6 $(+0.6\%)$ & 8.4 $(+13.5\%)$ & 5.2 $(+30\%)$ & 3.4 $(+41.7\%)$ & 13.1 $(+5.6\%)$ 	\\
PLM$_{PHP}$ & 16.8 $(+4.3\%)$ & 8.6 $(+8.9\%)$ & 5.1 $(+13.3\%)$ & 3.2 $(+14.3\%)$ & 12.8 $(+0\%)$ \\
PLM$_{Java}$ & 16.6 $(+3.1\%)$ & 8.8 $(+10\%)$ & 5.4 $(+17.4\%)$ & 3.5 $(+25\%)$ & 13.1 $(+3.1\%)$  \\
PLM$_{Python}$ & 15.9 $(+1.3\%)$ & 8.3 $(+10.7\%)$ & 5.1 $(+21.4\%)$ & 3.3 $(+26.9\%)$ & 13.1 $(+4.8\%)$  \\
PLM$_{Go}$ & 14.8 $(+0.7\%)$ & 7.8 $(+5.4\%)$ & 4.7 $(+11.9\%)$ & 3.2 $(+23.1\%)$ & 12.8 $(+1.6\%)$  \\ \hline
CodeBERT & 15.8 $(+2.6\%)$ & 8.7 $(+6.1\%)$ & 5.4 $(+12.9\%)$ & 3.5 $(+21.5\%)$ & 13.4 $(+1.5\%)$  \\
GraphCodeBERT & 16.6 $(+0.6\%)$ & 9.2 $(+1.1\%)$ & 5.8 $(+0\%)$ & 3.9 $(+0\%)$ & 13.5 $(+0\%)$  \\
\hline

\end{tabular}
\label{tab:rq5-code-summarization}
\end{center}
\end{table*}

\section{Threats to Validity}
\label{Section:Threats}

\textbf{External Validity.} 
%External validity refers to the ability to generalize results.
In this study, we discuss the results of different settings for Code Summarization and Code Search in Go, Java, Javascript, PHP, Python and Ruby.
% The WBO task is only available in Python. 
The tasks and the programming languages in our study are restricted, and the results might not be generalizable to other programming languages and tasks.
% Although the results might not be generalizable to other programming languages, especially the discussions of transferability of the learned knowledge by PLMs to programming languages, and the obtained results for Go language, the exploration of similar studies could benefit the research community.

\textbf{Internal Validity.} 
%This threat refers to the possibility of having unwanted or unanticipated relationships. 
% To alleviate the threats related to having unwanted results, we ensured the authors who trained the models are familiar with NLP and deep learning techniques. 
We process the publicly available datasets, following other research \cite{feng2020codebert, kanade2020learningCuBERT}. 
A potential threat may be related to not reaching the optimal performance of the pre-trained models, thus, having an under-trained PLM.
% them in the pre-training step. 
% This is due to the limited computational resources available for this research. 
We note that in the literature, there is no hard rule on the number of training steps for pre-training and to determine an optimal stopping criteria is still an open problem. Existing studies use fixed number of steps (a fraction of the dataset) as a stopping criteria \cite{bert,feng2020codebert,roberta}. For consistency, we pre-train all the PLMs involving different datasets for 50 epochs.
% To alleviate issues related to under training when comparing RoBERTa, we pre-trained all PLMs for 50 epochs, to make them consistent. 
% Additionally, instead of using the \textcolor{red}{CodeBERT?-base} model that is the RoBERTa model used in our study, we pre-trained RoBERTa from scratch. Therefore, mitigating the threat related to comparing the two architectures in our study. 
%other threats???: overfitting? Others?

\textbf{Construct Validity.} 
%Construction validity refers to the degree to which a test measures what it claims to be measuring. 
The validity threat here can be related to the measures used to evaluate the results. 
To mitigate the bias that might be related to a specific evaluation metric, we used multiple metrics that are commonly used in the downstream tasks. 
% These metrics are also used in related software engineering studies 
% \textcolor{red}{[REF]}
% . 
% \section{Discussion and Implication}
% \label{Section:Discussion-and-Implication}

\section{Conclusion and Future Works}
\label{Section:Conclusion-and-Future-Works}

In this work, we studied both monolingual and multilingual PLMs on Code Summarization and Code Search. We observed that the monolingual PLMs have better Performance-to-Time Ratio, as compared to the multilingual PLMs. In addition, our proposed strategies in selecting a suitable programming language for Code Summarization and Code Search are efficient and can improve the current state-of-the-art performance. Based on these findings, we suggest the following: 1) to consider using a monolingual PLM fine-tuned on a combined multilingual dataset as it has a higher Performance-to-Time Ratio, and 2) use our proposed strategies for Code Summarization and Code Search.
% One future direction of this research is the exploration of 
% low-resource languages, as the programming languages are different in many aspects including their syntax and static-dynamic typing.
% In the future, we aim to explore the ability and capacity of the multilingual PLMs when trained on many programming languages compared to a model trained on a single language. This can be done for multiple languages and tasks to explore other transfer learning techniques. 
In future, we aim to study the applicability of PLMs on more downstream tasks, and generalizing a strategy that can work well in multiple downstream tasks. 

\begin{acks}
This research is support by a grant from Natural Sciences and Engineering Research Council of Canada RGPIN-2019-05175.
\end{acks}

% \section*{Acknowledgment}
% This research is supported by a grant from GRANT-AGENCY-NAME. 

\break
\balance
\bibliographystyle{ACM-Reference-Format}
{\footnotesize\bibliography{paper}}

% that's all folks
\end{document}